\documentclass[prb,aps,twocolumn,showpacs]{revtex4}
\usepackage{amsfonts,amsmath,bm}
\usepackage[OT4]{fontenc}
\usepackage[T1]{fontenc}

\usepackage[pdftex]{graphicx}
\usepackage{epstopdf}

\newcommand{\rr}{{\bm{r}}}

\begin{document}

\title{Phonon-assisted relaxation between hole states in quantum dot molecules} 

\author{Krzysztof Gawarecki}

 \email{Krzysztof.Gawarecki@pwr.wroc.pl} 
\author{Pawe{\l} Machnikowski}
\affiliation{Institute of Physics, Wroc{\l}aw University of
Technology, 50-370 Wroc{\l}aw, Poland}

\begin{abstract}

We study theoretically phonon-assisted relaxation and inelastic tunneling of holes in a double quantum dot.
We derive hole states and relaxation rates from $\bm{k}\!\cdot\!\bm{p}$ Hamiltonians and show that there is a finite distance between the dots where 
lifetimes of hole states are very long which is related to vanishing tunnel coupling. We show also that the light hole admixture to hole states can  affect the hole relaxation rates even though its magnitude  is very small.
\end{abstract}


\maketitle


Systems composed of two stacked quantum dots are interesting due to their wide spectrum of non-trivial physical properties. In closely spaced double quantum dot structures, carrier states can be strongly delocalized \cite{bayer01} (like in a chemical covalent bond in molecules), hence these systems are often 
called quantum dot molecules (QDMs). They are promising for many possible applications, e.g., in quantum computing and nanoelectronics \cite{}.
Holes in QDMs can be particularly interesting due to their possible applications in quantum computing. 
Because of the long hole spin lifetime \cite{heiss07} and long coherence times, holes has been proposed as realization of qubit \cite{stinaff06,brunner09,hsieh09,doty08,doty10}. Hole states can be optically controlled using picoseconds optical pulses \cite{degreve11}. Furthermore, recent results show that hole spin states can be prepared with high fidelity \cite{godden10}. 
However, phonon-related processes are inevitable in a crystal environment and
may limit the feasibility of quantum control in these systems. 

The properties of the hole states are nontrivial as a result of subband mixing. In particular, recent theoretical predictions\cite{jaskolski04,jaskolski06,climente08a}  and experiments\cite{doty09} indicate that  
hole states in QDMs show an unusual behavior: the ground state of the hole
becomes antibonding above a certain critical distance between the dots. Thus, there is a distance where degeneracy between the bonding and antibonding molecular states occurs.
It can be expected that this behavior  will be reflected in the  phonon-assisted tunneling rates between the dots, leading to qualitatively
new behavior that was not present in the widely studied electron case \cite{gawarecki10,wu05,lopez05,stavrou05,vorojtsov05,climente06,grodecka08a,grodecka10}. These effects result from the non-trivial structure of hole states (eg. subband mixing) and cannot be revealed in simple models of hole states. Hole relaxation rates have been recently  measured \cite{wijesundara11} but only a simple model was used to interpret the results, where band mixing effects were not taken account.  

In this work, we present theoretical results for hole relaxation rates in a self-assembled QDM. In view of the important role of subband mixing for hole states in QDMs \cite{jaskolski04,jaskolski06,climente08a,doty09}, we propose an approach based on multi-band $\bm{k}\!\cdot\!\bm{p}$ method with the hole-phonon coupling described by the Bir-Pikus strain Hamiltonian. Our results show that for a certain finite distance (where the tunnel splitting vanishes), phonon-assisted relaxation is dramatically slowed down. We also prove that even small (below $2\%$) light hole admixture to hole states can strongly  affect the deformation potential coupling between carriers and phonons (up to $25\%$ of the corresponding relaxation rate).

The paper is organized as follows. First, we define
the model of the system under consideration. Next, we present the method used to calculate hole states. Subsequently, phonon-assisted relaxation of hole states is discussed. 

We consider an axially symmetric system formed by two geometrically identical self-assembled InAs dots in a GaAs matrix. The shape of both the dots is modeled as a spherical segment with the base radius of 10~nm and the height 3.7~nm.
Dots are placed on a wetting layer with thickness $0.6$~nm. The dots are separated by a distance $D$ (base to base) which is a variable parameter of our model. A 
diffusion layer of a very small thickness $0.3$~nm is included at the contact between the two materials,  where the material composition changes linearly. The system is placed in an axial electric field. The strain due to lattice mismatch is represented by the strain tensor $\hat \epsilon$ and calculated by minimizing the elastic energy of the system in the continuum elasticity approximation\cite{pryor98b}.

We use the 8-band Kane Hamiltonian with Bir-Pikus strain dependent terms \cite{pryor98b,andrzejewski10} which introduces dependence on position via the position dependent strain fields and parameter discontinuities at the InAs/GaAs border. We apply Burt-Foreman operator ordering \cite{foreman93,voon2009k}. 
Using the quasi degenerated L\"owdin perturbation method \cite{lowdin51,voon2009k} we reduce the problem to the 4x4 part describing the heavy hole (hh) and light hole (lh) subbands of the valence band.
For all the elements of this part we calculate the perturbative corrections resulting from the coupling to the conduction band up to order $k^{2}$. 
Corrections of order $k^2$ for diagonal elements are expressed in terms of (position dependent) effective masses. 
The corrections from the spin-orbit subbands are of the order $k^4$ and therefore are neglected. 

The matrix representation of the effective valence band Hamiltonian in the basis \{$|\mathrm{hh}\!\uparrow\rangle,|\mathrm{lh}\!\uparrow\rangle,|\mathrm{lh}\!\downarrow\rangle,|\mathrm{hh}\!\downarrow\rangle$\},
where $\uparrow$ and $\downarrow$ represent the projection of the total angular momentum ($\pm \frac{3}{2}$ for hh and $\pm \frac{1}{2}$ for lh), is
\begin{equation}
\label{matrixham}
H'=
\left (  
\begin{array}{cccc}
\tilde{P} & -\tilde{S} & \tilde{R} & 0 \\
-\tilde{S}^{\dagger} & \tilde{Q} & 0 & \tilde{R} \\
 \tilde{R}^{\dagger} & 0 & \tilde{Q} & \tilde{S} \\
 0 & \tilde{R}^{\dagger} & \tilde{S}^{\dagger} & \tilde{P}
\end{array}
\right ),
\end{equation}
where (in cylindrical coordinates)
\begin{align*}
\tilde{P} &= - k_{\bot} \frac{\hbar^2 }{2 m_{hh,\bot}} k_{\bot} - k_{z} \frac{\hbar^2 }{2 m_{hh,z}} k_{z} +E_{\mathrm{hh}}(\rho,z) + \varepsilon_{z} ,\\ 
\tilde{Q} &= - k_{\bot} \frac{\hbar^2 }{2 m_{lh,\bot}} k_{\bot} - k_{z} \frac{\hbar^2 }{2 m_{lh,z}} k_{z} +E_{\mathrm{lh}}(\rho,z) +  \varepsilon_{z} ,\\ 
\tilde{S} &= \sqrt{3} \frac{\hbar^2 }{m_{0}} ( k_{-} C_{1} k_{z} + k_{z} C_{2} k_{-} ) +d_{v} \epsilon_{r z} e^{-i \varphi},\\ 
\tilde{R} &= \sqrt{3} \frac{\hbar^2}{2 m_{0}} k_{-} C_{3} k_{-} +\frac{\sqrt{3}}{2} b_{v} (\epsilon_{r r} - \epsilon_{\varphi \varphi } ) e^{-2 i \varphi},
\end{align*} 
and
\begin{align*}
C_{1} &= - 1 - \gamma'_{1} + 2 \gamma'_{2} + 6 \gamma'_{3} -  \frac{E_{p}}{2  E_{g,hh}(\rho,z)}  - \frac{E_{p}}{ 2 E_{g,lh}(\rho,z)}   ,\\ 
C_{2} &= 1 + \gamma'_{1} -2 \gamma'_{2} ,\\ 
C_{3} &= \gamma'_{2} + \gamma'_{3} + \frac{1}{6} \left ( \frac{E_{p}}{ E_{g,hh}(\rho,z)}  + \frac{E_{p}}{E_{g,lh}(\rho,z)}  \right ).\\ 
\end{align*} 
Here $ \varepsilon_{z}$ is an electric field, $\gamma'_{i}$ are modified Luttinger parameters~\cite{boujdaria01},
the operators  $k_{\pm}$ in cylindrical coordinates are given by 
\begin{displaymath}
k_{\pm} = e^{\pm i \varphi} \left ( \pm \frac{1}{r} \frac{\partial}{\partial \varphi} - i \frac{\partial }{ \partial r} \right ),
\end{displaymath}
$m_{0}$ is the free electron mass,
effective masses are defined as
\begin{align*}
m^{-1}_{\mathrm{hh},z}(\rho, z) &= \left ( \gamma'_{1} - 2 \gamma'_{2} \right ) m^{-1}_{0}, \\
m^{-1}_{\mathrm{hh},\bot}(\rho, z) &=  \left (\gamma'_{1} + \gamma'_{2} + \frac{E_{p}}{2 E_{g,hh}(\rho,z)}\right ) m^{-1}_{0}, \\
m^{-1}_{\mathrm{lh},z}(\rho, z) &=  \left ( \gamma'_{1} + 2 \gamma'_{2} +  \frac{2 E_{p}}{3 E_{g,lh}(\rho,z)}\right ) m^{-1}_{0}, \\
m^{-1}_{\mathrm{lh},\bot}(\rho, z) &=  \left ( \gamma'_{1} - \gamma'_{2} + \frac{E_{p}}{6 E_{g,lh}(\rho,z)}\right ) m^{-1}_{0},
\end{align*} 
where 
\begin{align*}
E_{\mathrm{g,hh}} (\rho,z) & =  E_{\mathrm{c}} (\rho,z)  - E_{\mathrm{hh}} (\rho,z), \\
E_{\mathrm{g,lh}} (\rho,z) & =  E_{\mathrm{c}} (\rho,z)  - E_{\mathrm{lh}} (\rho,z).
\end{align*}
Position-dependent band edges $E_{\mathrm{c}}(\rho,z)$, $E_{\mathrm{hh}}(\rho,z)$ and $E_{\mathrm{lh}}(\rho,z)$ are  given by
\begin{align*}
E_{\mathrm{c}} (\rho,z) & =  E_{\mathrm{c}0}+a_{c} \mathrm{Tr} \{ \hat \epsilon \}, \\
E_{\mathrm{hh}} (\rho,z) & =  E_{\mathrm{v}0}-a_{v} \mathrm{Tr} \{ \hat \epsilon \}- b_{v} [\epsilon_{zz} - 0.5(\epsilon_{xx} +\epsilon_{yy} )], \\
E_{\mathrm{lh}} (\rho,z) & =  E_{\mathrm{v}0}-a_{v} \mathrm{Tr} \{ \hat \epsilon \} + b_{v} [\epsilon_{zz} - 0.5(\epsilon_{xx} +\epsilon_{yy} )],
\end{align*}
where $a_{\mathrm{\mathrm{v}}}$ and $b_{\mathrm{v}}$ are the valence band deformation
potentials and $a_{\mathrm{\mathrm{c}}}$ is the conduction band deformation potential. The values of the material parameters are listed in Table~\ref{tab:param}\cite{pryor98b}. 
The band offsets assumed here are close to the values in Ref. \onlinecite{vurgaftman01}.
\begin{table}
\begin{tabular}{llll}
\hline
& & GaAs & InAs \\
\hline
Band structure parameters & $E_{\mathrm{c}0}$ & 0.95~eV & 0 \\
 & $E_{\mathrm{v}0}$ & -0.57~eV & -0.42~eV \\
 & $E_{\mathrm{p}}$ & 25.7~eV & 22.2~eV \\
Modified Luttinger parameters & $\gamma'_{1}$ & 1.34 & 1.98 \\
 & $\gamma'_{2}$ & -0.57 & -0.44 \\
 & $\gamma'_{3}$ & 0.062 & 0.48  \\
Deformation potentials & $a_{\mathrm{c}}$ & -9.3~eV & -6.66~eV \\
 & $a_{\mathrm{v}}$ & 0.7~eV & 0.66~eV \\
 & $b$ & -2.0~eV & -1.8~eV \\
Speed of sound - longitudinal& $c_{\mathrm{l}}$ & \multicolumn{2}{c}{5150~m/s} \\
\hspace{21.5 mm} - transverse& $c_{\mathrm{t}}$ & \multicolumn{2}{c}{2800~m/s} \\
Crystal density & $\varrho$ & \multicolumn{2}{c}{5300~kg/m$^{3}$} \\
Piezoelectric constant & $d$ &
\multicolumn{2}{c}{-0.16~C/m$^{2}$} \\
Relative dielectric constant & $\varepsilon_{\mathrm{r}}$ &
\multicolumn{2}{c}{12.9} \\ 
\hline
\end{tabular}
\caption{\label{tab:param}Material parameters used in the calculations.}
\end{table}

The Hamiltonian describing the hole-phonon coupling is \cite{woods04}
\begin{align}
\label{ham:tun}
H_{\mathrm{int}} &  = \sum_{n n'} \langle \psi_{n} | H_{\mathrm{B-P}} + H_{\mathrm{PE}}| \psi_{n'} \rangle a_{n}^{\dagger} a_{n'} \nonumber \\ &= 
\sum_{n n'} \int d^{3} \rr  \psi^{*}_{n} (\rr) ( H_{\mathrm{B-P}} + \hat{I} V(\rr)  )\psi_{n'} (\rr)  a^{\dagger}_{n} a_{n'},
\end{align}
where $\psi_{n}$ is the four-component eigenfunction the elements of which are related to the valence subbands, $H_{\mathrm{B-P}}$ is the valence part (4x4) of the Bir-Pikus Hamiltonian describing the deformation potential (DP) coupling, $H_{\mathrm{PE}}=\hat{I} V(\rr)$ is the part of the Hamiltonian describing piezoelectric (PE) coupling and $\hat{I}$ is the 4x4 identity matrix.
The piezoelectric potential is given by 
  $V(\rr) = -i (\hat{d} \cdot \hat{\epsilon})_{\lVert} / ( \varepsilon_{0} \varepsilon_{r} )$
where $\hat{d}$ is the piezoelectric tensor,  $\varepsilon_{0}$ is the vacuum permittivity and $\varepsilon_{r}$ is the relative dielectric constant of GaAs. In the zincblende structure, the only non-zero components of $\hat{d}$ are
$d_{x y z}=d_{y z x}=d_{z x y}=d$. 
Because of the orthogonality and identical parity of Bloch functions for different valence subbands, the lowest order inter-subband contribution from the piezoelectric coupling contains an additional factor on the order of $(qa)^2$, where $a$ is the lattice constant. This term not only qualitatively affects the small q behavior of the coupling constant but also is quantitatively small for low values of q. Therefore, for long wavelength phonons, the inter-subband elements of the piezoelectric potential are much smaller (by 2 orders of magnitude in GaAs) than those resulting from the DP coupling.
 We take into account only acoustic phonons because the inter-level energy distance in our structure is much smaller than the optical phonon energy. We assume linear dispersion of acoustic phonons. 


In order to describe the hole states and account for the subband mixing effects we propose a multi-subband generalization of the `adiabatic' separation of variables \cite{wojs96,gawarecki10}. 
For each subband, we solve numerically the one-dimensional equation along the growth ($z$) direction for each value of the radial coordinate $\rho$,
\begin{eqnarray*} 
\left[  -\frac{\partial}{\partial z}
\frac{\hbar^{2}}{2m_{\alpha,z}(\rho,z)} \frac{\partial}{\partial z}
+E_{\mathrm{\alpha}}(\rho,z)\right]\chi_{\alpha}(\rho,z) \\=\varepsilon_{\alpha}(\rho)\chi_{\alpha}(\rho,z),
\end{eqnarray*}
where $\alpha$ is a subband index. We find the two lowest solutions to this equation for the hh and lh subbands and obtain $\chi_{\mathrm{hh},n}(\rho,z)$,$\chi_{\mathrm{lh},n}(\rho,z)$, $n=0,1$ as well as the corresponding $\rho$-dependent eigenvalues which can be interpreted as effective potentials for the radial problem. In the next step, we apply the Ritz variational method, looking
for the stationary points of the functional 
$F[\psi]=\langle \psi|H|\psi\rangle$.
We use the class of normalized trial functions 
\begin{align*}\label{ansatz}
\psi(\rho,z,\phi)=&\frac{1}{\sqrt{2\pi}}\sum_{n} 
 \begin{pmatrix}\chi_{\mathrm{hh} ,n}(\rho,z)\varphi_{\mathrm{hh} \uparrow,n}(\rho)e^{i(M-3/2)\phi}\\\chi_{\mathrm{lh} ,n}(\rho,z)\varphi_{\mathrm{lh} \uparrow,n}(\rho)e^{i(M-1/2)\phi}\\\chi_{\mathrm{lh} ,n}(\rho,z)\varphi_{\mathrm{lh} \downarrow,n}(\rho)e^{i(M+1/2)\phi}\\\chi_{\mathrm{hh} ,n}(\rho,z)\varphi_{\mathrm{hh} \downarrow,n}(\rho)e^{i(M+3/2)\phi}\end{pmatrix},
\end{align*}
where $M$ is the projection of the angular momentum on the system axis and the four components refer to the four valence subbands. Finally, we numerically minimize the functional $F[\psi]$ on a grid and obtain the two lowest eigenfunctions of the system.

\begin{figure}
\begin{center}
\includegraphics[width=85mm]{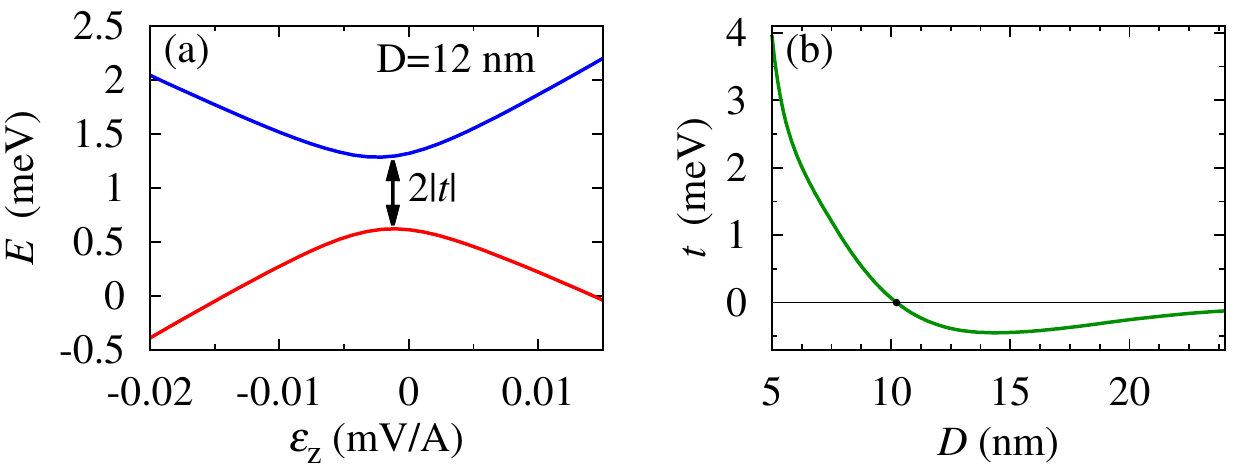}
\end{center}
\caption{\label{fig:tparam}(Color online) (a) Energy of the two lowest hole states for $D=12$~nm. The absolute value of the tunnel coupling parameter is equal to half of the energy splitting at the resonance. (b) The tunnel coupling parameter as a function of the distance between the dots. The point of $t=0$ is where the bonding--antibonding transition of the ground state takes place.}
\end{figure}

The obtained lowest eigenstates have a small lh admixture (from 0.5\% to 2\% contribution to the total state) due to a very weak confinement of light holes in our structure. In order to characterize the tunnel coupling between the hole states in the two dots, we study the anticrossing between the two lowest states in the QDM as the electric field is scanned through resonance. The effective tunnel coupling parameter is defined as half of the anticrossing width, with a positive (negative) sign assigned in the case of bonding (antibonding) ground state. The tunnel coupling parameter $t$ is shown in Fig.~\ref{fig:tparam}(b), as a function of the distance between the dots $D$. The transition from bonding to antibonding ground state, which occurs about $D=10.23$~nm in our structure, corresponds to vanishing tunnel coupling between the dots.



\begin{figure}
\begin{center}
\includegraphics[width=90mm]{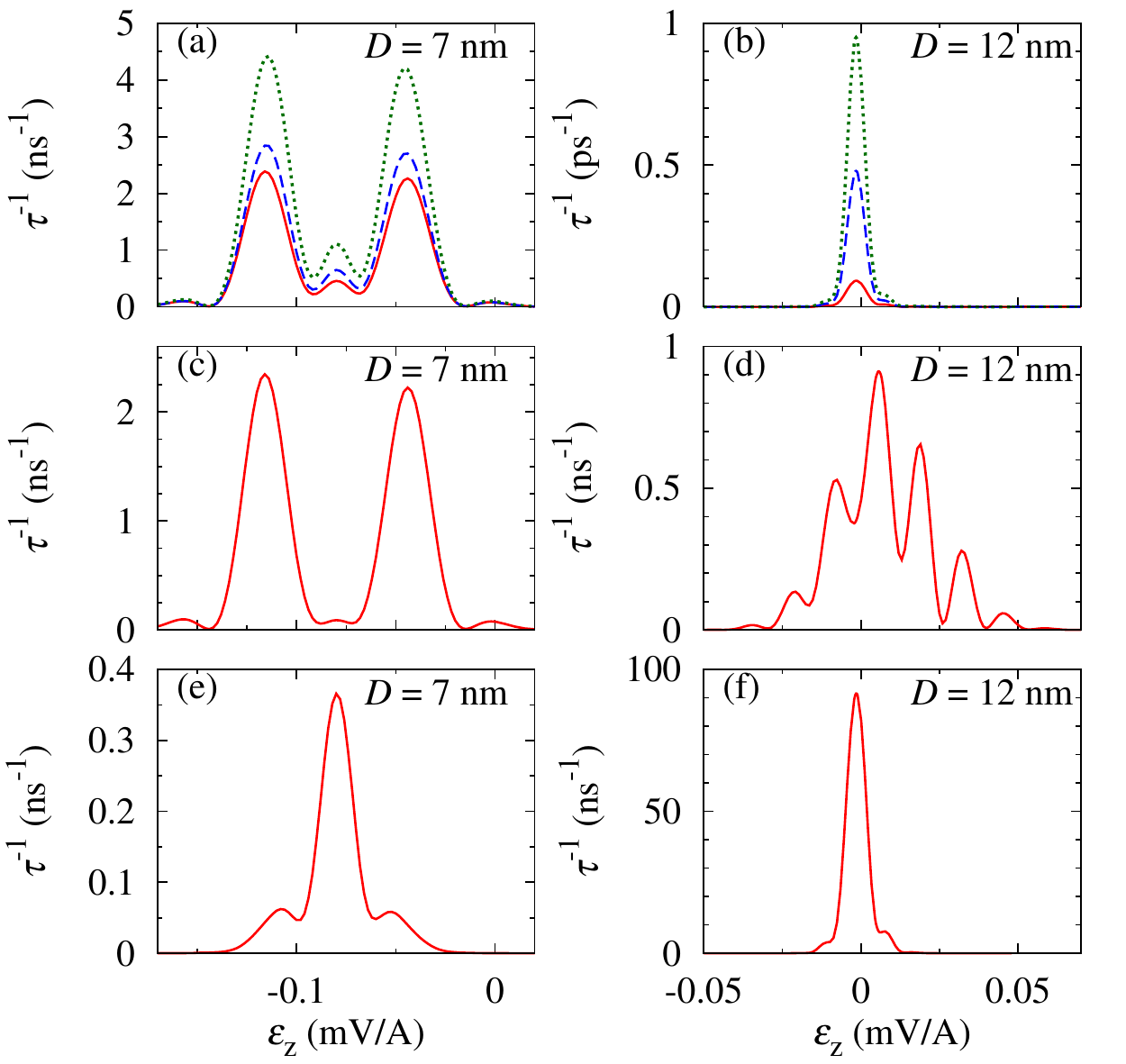}
\end{center}
\caption{\label{fig:relaxation}(Color online) (a-b)
Total phonon assisted relaxation rate for two different distances between dots  at $T=0$~K (red solid
line), 20~K (blue dashed line), and 40~K (green dotted line).
(c,d) Contributions to the phonon-induced relaxation
rate due to DP coupling.  (e,f) Contributions from PE coupling.}
\end{figure}

We calculate phonon-assisted relaxation rates using the Fermi golden rule\cite{grodecka08a} with the Hamiltonian (\ref{ham:tun}).
The total phonon-assisted relaxation rate for two different distances between the dots ($D=7$~nm and $12$~nm) are shown in Fig.~\ref{fig:relaxation}(a,b). The rates are calculated as a function of the axial electric field at three temperatures ($T=0$, $20$~K and $40$~K). The contributions from the DP and PE couplings to the total relaxation rate are shown in the Fig.~\ref{fig:relaxation}(c-f). For closely spaced dots ($D=7$~nm) the energy splitting $2t$ is high. Then, relaxation rates from DP and PE couplings are comparable, although relaxation rate from DP slightly dominates.
On the contrary, for larger distances, the PE interaction is much stronger. This behavior is similar to the electron case \cite{gawarecki10}. The DP coupling dominates for large energy splittings because this coupling is isotropic and involves LA phonons which have higher energies, while the PE coupling is anisotropic and is suppressed for emission along the $z$ direction which is preferred at high energies \cite{gawarecki10}. On the other hand, for small splittings, the PE coupling becomes realatively stronger because of its strong $1/q$ dependence at $q \rightarrow 0$.

\begin{figure}
\begin{center}
\includegraphics[width=85mm]{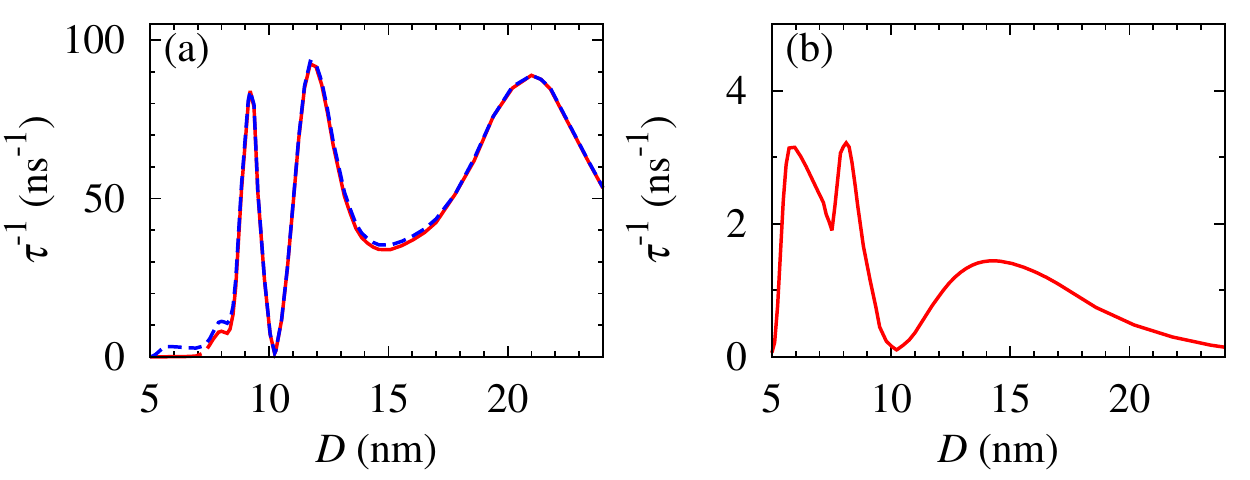}
\end{center}
\caption{\label{fig:tunmax}(Color online) (a) The total maximum relaxation rate (blue dashed line) and the contribution from the PE coupling (red solid line). (b) Contribution to the total relaxation rate from the DP coupling.}
\end{figure}

In order to study the overall dependence of the relaxation rate on the inter-dot separation, we calculate the maximum (over field magnitudes) phonon-assisted relaxation rate as a function of $D$. The total maximum relaxation rate as well as the
contributions from the DP and PE couplings  are shown in Fig.~\ref{fig:tunmax}(a-b). For $D \rightarrow 0$ both the relaxation rates drop down because the energy splitting is large and the density of phonon states at very high frequencies is low. On the other hand, for $D \rightarrow \infty$ both the relaxation rates decay due to vanishing overlap between the wavefunctions. One can see that for distances near $D= 10.23$~nm (where $t=0$) the maximum phonon-assisted relaxation rate drops down to $\tau^{-1} = 0.0798$~ns$^{-1}$ and $\tau^{-1} = 1.04$~ns$^{-1}$ for the PE and DP couplings respectively, that is, by two orders of magnitude compared to the highest values. This results from an extremely narrow energy splitting in view of low density of states of low frequency phonons. Clearly, the relaxation at the critical distance remains very slow at any electric field.

\begin{figure}
\begin{center}
\includegraphics[width=88mm]{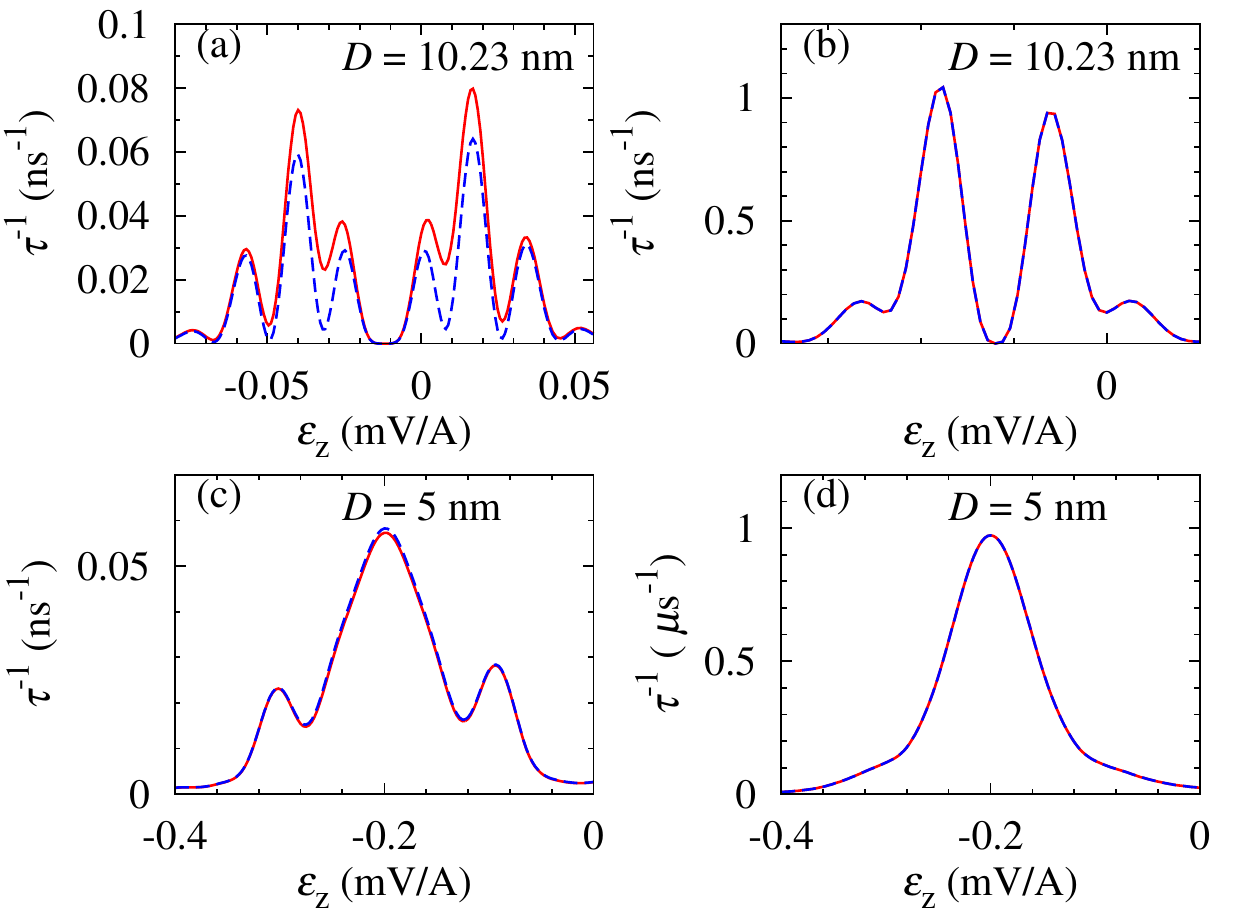}
\end{center}
\caption{\label{fig:comparation}(Color online) (a,c) Contribution to the relaxation rate due to the DP coupling. Results from 4-band model (red solid line) and from the single band approximation (blue dashed line). (b,d) The same comparison for the PE coupling.}
\end{figure}
We have also verified the accuracy of the single band approximation. We have made calculation of the phonon assisted relaxation rate for $D=10.23$~nm (Fig.~\ref{fig:comparation}(a,b)) and for $D=5$~nm (Fig.~\ref{fig:comparation}(c,d)) and compared the results for the DP  (Fig.~\ref{fig:comparation}(a,c))  and PE couplings (Fig.~\ref{fig:comparation}(b,d)). For the DP coupling at $D=10.23$~nm, the difference between the full model and the single band approximation is considerable. This can be interpreted as follows. Near the point $t=0$, the two lowest eigenfunctions are strongly localized on different dots and the overlap of their hh components is very small. However, lh components of these wavefunctions are still delocalized due to the shallow confinement in the lh subband. This leads to a strong coupling between hh component of the first state and the lh component of the second one via the inter-subband elements of the strain (Bir-Pikus) Hamiltonian. On the other hand, for PE interaction, the mismatch between the models is very small. The reason is the absence of inter-subband terms for PE coupling. We can also see that for both the DP and PE cases the relaxation rate vanishes at a certain electric field corresponding to the exact resonance. At $D=5$~nm (Fig.~\ref{fig:comparation}c) for DP coupling, the discrepancy between the two models is much smaller than for  $D=10.23$~nm. 
For stronger tunnel coupling (smaller $D$), the resonance area, where the wavefunctions are delocalized, extends over a wide range of electric fields. 
In consequence, the interband coupling is not so important in this case.  For even larger values of $D$, localization is strong, but lh contribution to the states decreases, hence the contribution from the interband coupling is also small. 

In summary we have studied phonon-assisted relaxation between the two lowest hole states in a QDM, focusing on the role of subband mixing and the bonding-antibonding transition of the ground state. We have shown that the relaxation rate is very small at the critical distance (degeneracy point), where the character of the ground state wavefunction changes. Our findings are consistent with the general features of
experimental observations \cite{wijesundara11}, where hole lifetimes on the order of a few nanoseconds were observed (the geometry of the system studied in that work was, however, different from that considered here). We have also investigated the influence of the lh admixture on the relaxation rate. We have shown that near the degeneracy point, the subband mixing gives an important correction to phonon-induced relaxation via DP channel, up to $25\%$ in some cases, in spite of a very small contribution of the lh subband to the hole state (below $2\%$). Our results show that phonon-assisted hole relaxation and tunneling in QDMs are with some respects qualitatively different from the electron case. The complete description of the hole-phonon kinetics turns out to be impossible without allowing for subband mixing in contrast to the electron case, where simple models are able to essentially correctly account for the relaxation rates.

We are grateful to Anna Musia{\l} for asking inspiring questions.
This work was supported by the
Foundation for Polish Science under the TEAM programme,
co-financed by the European Regional Development Fund.

\bibliographystyle{prsty}
\bibliography{abbr,quantum}
\end{document}